\documentclass{article}
\usepackage{ijcai26}
\usepackage{authblk}
\usepackage{times}
\usepackage{soul}
\usepackage{url}
\usepackage[hidelinks]{hyperref}
\usepackage[utf8]{inputenc}
\usepackage[small]{caption}
\usepackage{graphicx}
\usepackage{amsmath}
\usepackage{amsthm}
\usepackage{booktabs}
\usepackage{algorithm}
\usepackage{algorithmic}
\usepackage[switch]{lineno}
\usepackage{amssymb}
\usepackage{multirow}

\title{BioLM-Score: Language-Prior Conditioned Probabilistic Geometric Potentials for Protein-Ligand Scoring}

\begin{document}

\author[1,$\dagger$]{\textbf{Zhangfan Yang}}
\author[2,$\dagger$]{\textbf{Baoyun Chen}}
\author[2]{\textbf{Dong Xu}}
\author[2]{\textbf{Jia Wang}}
\author[1]{\textbf{Ruibin Bai}}
\author[2]{\textbf{Junkai Ji}}
\author[2,*]{\textbf{Zexuan Zhu}}

\affil[1]{School of Computer Science, University of Nottingham Ningbo, Ningbo, China}
\affil[2]{School of Artificial Intelligence, Shenzhen University, Shenzhen, China}
\affil[$\dagger$]{\small Equal contribution}
\affil[*]{\small Corresponding author: \texttt{zhuzx@szu.edu.cn}}
\date{}

\maketitle

\begin{abstract}
Protein-ligand scoring is a central component of structure-based drug design, underpinning molecular docking, virtual screening, and pose optimization. Conventional physics-based energy functions are often computationally expensive, limiting their utility in large-scale screening. In contrast, deep learning-based scoring models offer improved computational efficiency but frequently suffer from limited cross-target generalization and poor interpretability, which restrict their practical applicability. Here we present BioLM-Score, a simple yet generalizable protein-ligand scoring model that couples geometric modeling with representation learning.
Specifically, it employs modality-specific and structure-aware encoders for proteins and ligands, each augmented with biomolecular language models to enrich structural and chemical representations. Subsequently, these representations are integrated through a mixture density network to predict multimodal interatomic distance distributions, from which statistically grounded likelihood-based scores are derived. Evaluations on the CASF-2016 benchmark demonstrate that BioLM-Score achieves significant improvements across docking, scoring, ranking, and screening tasks. Moreover, the proposed scoring function serves as an effective optimization objective for guiding docking protocols and conformational search. In summary, BioLM-Score provides a principled and practical alternative to existing scoring functions, combining efficiency, generalization, and interpretability for structure-based drug discovery.
\end{abstract}

\section{Introduction}

Structure-based drug design (SBDD) is pivotal to drug repurposing strategies~\cite{mayr2009novel}. Given the prohibitive costs and high attrition rates of \textit{de novo} drug discovery, repurposing approved drugs for novel targets has emerged as a cost-effective alternative~\cite{ban2017best,drews2000drug}. Structure-based virtual screening is a primary method for identifying such candidates~\cite{gorgulla2020open}. It relies on scoring functions to evaluate protein-ligand interaction modes, where the highest-ranked pose is hypothesized to be the most native-like~\cite{song2021protein}. Despite decades of advancement, developing a scoring function that effectively balances computational efficiency, cross-target generalizability, and interpretability remains a formidable challenge in the field~\cite{su2018comparative}.

Traditional scoring functions are generally categorized into physics-based and empirical methods~\cite{yang2022protein}\cite{aldeghi2016accurate}. While physics-based approaches offer theoretical rigor through explicit force fields, they incur prohibitive computational costs~\cite{monticelli2012force}. Conversely, empirical functions, though efficient, are constrained by rigid functional forms and linear assumptions, which limit their capacity to capture complex non-linear interactions~\cite{wu2018coach,meli2022scoring}. Recently, data-driven deep learning (DL) approaches have emerged as promising alternatives, leveraging vast structural datasets to capture complex non-linear interactions~\cite{chauhan2018review,jin2023capla}. However, standard DL models formulating scoring as direct affinity regression often struggle with out-of-distribution generalization to novel targets and lack the interpretability essential for rational drug design~\cite{shen2021beware}.

A promising recent direction focuses on learning geometric likelihoods directly from native complex structures. Methods such as mixture density networks (MDNs) estimate interatomic distance distributions to parameterize statistical potentials, ensuring strong geometric consistency~\cite{zhang2024advancing,xia2025normalized}. Nevertheless, two critical limitations persist. First, geometric plausibility is not equivalent to binding affinity. Likelihood-based scores are optimized to maximize the probability of native-like conformations, which does not necessarily correlate with binding affinity, often leading to poor ranking performance in affinity-driven tasks~\cite{zhang2025graph,luo2024enhancing}. Second, current geometric approaches predominantly focus on local 3D graph environments, thereby overlooking rich global contexts, such as evolutionary information encoded in protein sequences and chemical semantics within ligand SMILES strings~\cite{lam2024protein,chen2025hitscreen}.

To address these limitations, we present BioLM-Score, a generalizable scoring framework that unifies geometric modeling with representation learning. Unlike purely graph-based approaches, BioLM-Score integrates modality-specific geometric encoders with pre-trained biomolecular language models (BioLMs). This dual-branch architecture allows the model to condition local geometric interactions on global chemical and evolutionary priors derived from protein sequences and ligand SMILES. Specifically, these enriched representations are fused to parameterize a mixture density network (MDN), which predicts multimodal interatomic distance distributions. The final score is computed as a probabilistic log-likelihood, facilitating both physical plausibility and interpretability.

We comprehensively evaluate BioLM-Score on the standard CASF-2016~\cite{su2018comparative} and DEKOIS 2.0~\cite{bauer2013evaluation}. Empirical results demonstrate that our model achieves state-of-the-art performance across scoring, ranking, docking, and screening tasks. Furthermore, we demonstrate that BioLM-Score serves as a robust objective function for guiding conformational search within evolutionary docking protocols. By effectively bridging geometric deep learning with the semantic reasoning of language models, BioLM-Score offers a principled and practical solution for high-throughput structure-based drug discovery.

Our contributions can be summarized as follows: 
\begin{itemize} 
    \item We introduce a novel protein–ligand scoring framework that systematically integrates graph neural networks with biomolecular language models, enabling unified structure-aware modeling and biomolecular representation learning within a single framework.
    \item We empirically demonstrate substantial performance gains on the CASF-2016 benchmark, and further establish that the proposed scoring function can serve as an effective standalone optimization objective, leading to improved accuracy in conformational search and binding pose identification.
\end{itemize}

\begin{figure*}[htbp] 
    \centering
    \includegraphics[width=1.0\linewidth]{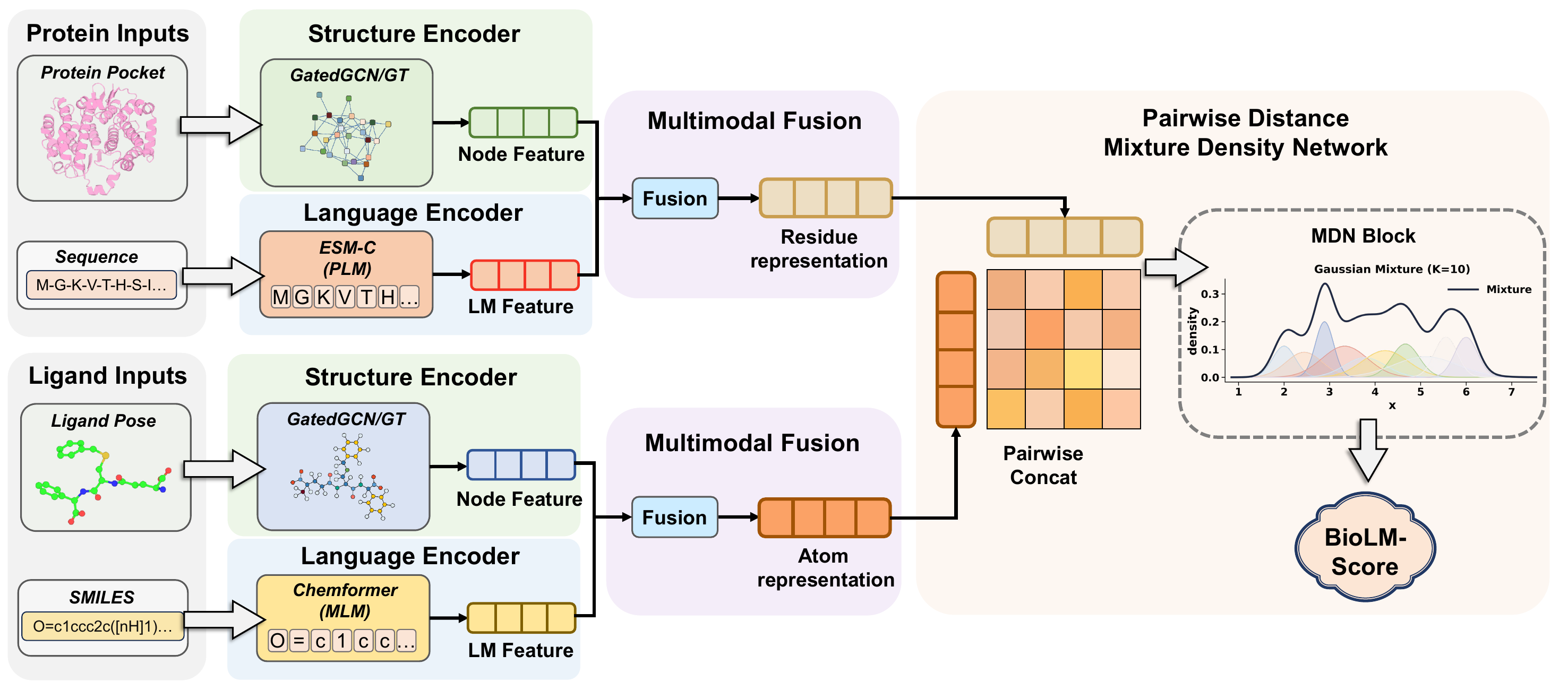} 
    
    \caption{The overall architecture of BioLM-Score. The model processes protein and ligand inputs through dual encoders: a Structure Encoder (GatedGCN/GT) and a Language Encoder (ESM-C for proteins, Chemformer for ligands) The extracted features undergo multimodal fusion to generate residue and atom features. Finally, a Pairwise Distance Mixture Density Network (MDN)  predicts the interaction density to compute the final BioLM-Score.}
    
    \label{fig:BioLM_architecture}
\end{figure*}

\section{Related Work}

Protein-ligand scoring functions are pivotal for virtual screening. Existing approaches are categorized into physics-based and empirical methods, data-driven machine learning (ML), and likelihood-based paradigms.

\paragraph{Physics-based and Empirical Methods}
Classical tools, including X-score~\cite{x-score}, AutoDock~\cite{morris2009autodock4}, AutoDock Vina~\cite{vina}, Glide~\cite{glide}, and GOLD~\cite{gold}, approximate binding free energy via linear combinations of physical terms (e.g., van der Waals) and empirical descriptors. While these methods enforce geometric constraints, they are often limited to local feasibility and lack transferability across diverse targets, resulting in ranking biases during large-scale screening.

\paragraph{Data-Driven Machine Learning-based Scoring}
Early ML approaches, such as $\Delta_{vina}RF$~\cite{vinarf},  $\Delta_{vina}RF_{XGB}$~\cite{vinaxgb}, and $\Delta$-AEScore~\cite{deltaae}, operated as re-scoring functions to refine traditional energy terms. Subsequent methods like OnionNet-SFCT~\cite{onionnet} integrated interaction fingerprints to improve ranking. More recent architectures, including PIGNet~\cite{pignet},  PIGNet2~\cite{pignet2} and IGModel~\cite{igmodel}, adopt a discriminative paradigm, leveraging data augmentation to distinguish natives from decoys. However, relying on synthetic negatives conditions these models on sampling artifacts rather than physics. Consequently, they diverge from approximating a transferable potential energy surface (PES), effectively overfitting to the specific decoy selection protocols.

\paragraph{Geometric Likelihood Estimation}
To model interaction distributions explicitly, methods like DeepDock~\cite{deepdock} and RTMScore~\cite{rtmscore} utilize Mixture Density Networks (MDNs) and Graph Neural Network (GNN) to estimate geometric likelihoods. GenScore~\cite{genscore} further imposes affinity supervision upon this framework to align geometric probability with binding data. Despite these advances, current likelihood models rely predominantly on 3D graphs, overlooking the critical evolutionary and chemical semantics encoded in 1D protein-ligand sequences.

\paragraph{Proposed Approach.}
To address this limitation, we integrate protein and molecular language models with a geometric learning framework. By fusing sequence-based semantic priors with structure-based interaction modeling, our method injects complementary semantic cues into geometric representations, thereby enhancing scoring reliability.

\section{Method}
We propose BioLM-Score, a scoring function to evaluate the geometric plausibility of protein-ligand binding poses. Given a protein pocket $P$ and a ligand pose $L$, a real-valued scoring function $s(P,L)\in\mathbb{R}$ is learned to quantify their structural compatibility. The model follows a dual branch design. Structural features are extracted from protein and ligand graphs via a GNN backbone, while sequence priors are derived from language models pre-trained on protein sequences and ligand SMILES strings, respectively. To decouple the scoring formulation from specific backbone architectures, the structural branch implies interchangeable graph encoders, such as GatedGCN and Graph Transformers. Following node-level multimodal fusion, a MDN models pairwise residue-atom distance distributions. We formulate the final score as the aggregated log-likelihood of pairwise distances, effectively capturing the probabilistic geometric potential.

\subsection{Graph Construction and Feature Extraction}
\paragraph{Ligand graph.}We represent the ligand as an undirected graph $G_L=(V_L, E_L)$, where $V_L$ is the set of atoms and $E_L$ represents the covalent bonds. The node and edge features are detailed in {Supplementary Table~4}.
\paragraph{Protein pocket graph.}We represent the protein pocket as a residue-level undirected graph $G_P=(V_P, E_P)$, where $V_P$ denotes the set of pocket residues. An edge $(u,v) \in E_P$ is defined if the minimum inter-residue atomic distance between $u$ and $v$ is less than $10\,\text{\AA}$. Corresponding node and edge features are summarized in {Supplementary Table~5}.

\subsection{Structural Graph Encoders}
Let $\mathbf{x}_i$ and $\mathbf{a}_{ij}$ denote the initial node and edge feature vectors, respectively. Regardless of the backbone architecture, we first project these inputs into continuous embeddings via learnable linear transformations. To iteratively update these representations over $L$ layers, we investigate two distinct graph encoding architectures (detailed formulations are provided in {Supplementary~A}).

\paragraph{GatedGCN.}
As our primary backbone, we employ a Gated Graph Convolutional Network~\cite{bresson2017residual}. This architecture introduces an edge gating mechanism to modulate information flow during message passing. Specifically, it derives gating coefficients from both node and edge features, allowing the model to dynamically weigh the importance of neighbors during aggregation while simultaneously updating edge features via residual connections.

\paragraph{Structure-aware Graph Transformer (GT).} Alternatively, to improve information routing within local neighborhoods, we employ a structure-aware GT~\cite{dwivedi2020generalization}. Unlike simple gating, GT computes attention weights via query-key interactions explicitly modulated by edge attributes. Furthermore, the attention outputs are utilized to refine edge features, facilitating a joint optimization of node-edge representations.

\subsection{Sequence-based Encoding via Pre-trained Language Models}
Given the protein sequence $S_P$ and ligand SMILES string $S_L$, we employ ESM-C~\cite{esm2024cambrian} and Chemformer~\cite{irwin2022chemformer} to generate their respective token-level representations. The resulting embeddings are mapped to graph nodes (residues or atoms) and projected to the hidden dimension $d$ via independent MLPs:
\begin{equation}
\begin{aligned}
\bar{\mathbf{U}}_P &= \mathrm{MLP}_P\big( \mathrm{ESM\text{-}C}(S_P)[\mathcal{I}_P] \big), \\
\bar{\mathbf{U}}_L[v] &= \mathrm{MLP}_L\big( \mathrm{Chemformer}(S_L)[r_v] \big), \quad \forall v \in V_L.
\end{aligned}
\end{equation}
Here, $\mathcal{I}_P$ denotes the index set of pocket residues, and $r_v$ represents the token index corresponding to atom $v$ in the SMILES string.

\subsection{Node-Level Multimodal Fusion Variants}
To effectively integrate structural representations $\mathbf{h}$ (from graph encoder) and sequence priors $\mathbf{u}$ (from Protein Language Model/Molecular Language Model), we explore four distinct fusion strategies $\mathcal{F}(\mathbf{h}, \mathbf{u})$. Let $\mathbf{z}$ denote the fused embedding for a given node (residue or atom):
\begin{itemize}
  \item \textbf{Concatenation.} We concatenate the two feature vectors and project them via a learnable MLP, as described in the baseline setting:
\begin{equation}
    \mathbf{z}_{\text{cat}} = \mathrm{MLP}\left( [\mathbf{h}\, |\, \mathbf{u}] \right).
\end{equation}
    \item \textbf{Element-wise Summation.}
Assuming the feature spaces are aligned, we perform direct element-wise addition to superimpose sequence information onto structural features:
\begin{equation}
    \mathbf{z}_{\text{sum}} = \mathbf{h} + \mathbf{u}.
\end{equation}
  \item \textbf{Gated Fusion.}
A learnable gating mechanism dynamically weighs the contribution of each modality. We compute a gate coefficient $\mathbf{g} \in [0, 1]^d$ and perform a soft selection:
\begin{equation}
    \mathbf{g} = \sigma\left( \mathbf{W}g [\mathbf{h} \,|\, \mathbf{u}] \right), \quad \mathbf{z}_{\text{gate}} = \mathbf{g} \odot \mathbf{h} + (1 - \mathbf{g}) \odot \mathbf{u},
\end{equation}where $\sigma$ is the sigmoid function and $\mathbf{W}_g$ is a learnable weight matrix.
  \item \textbf{Cross-Attention.}
We employ a Multi-Head Cross-Attention (MCA) mechanism where the structural embedding serves as the query ($\mathbf{Q}$), and the sequence embedding serves as the key ($\mathbf{K}$) and value ($\mathbf{V}$):
\begin{equation}
    \mathbf{z}_{\text{attn}} = \mathbf{h} + \mathrm{Dropout}\left( \mathrm{MCA}(\mathbf{Q}=\mathbf{h}, \mathbf{K}=\mathbf{u}, \mathbf{V}=\mathbf{u}) \right).
\end{equation}
\end{itemize}
Following the fusion step, the resulting embeddings $\mathbf{Z}_P$ and $\mathbf{Z}_L$ are utilized for downstream interaction modeling.

\begin{table*}[t]
\centering
\small
\renewcommand{\arraystretch}{0.9}
\setlength{\tabcolsep}{5pt}
\resizebox{\linewidth}{!}{%
\begin{tabular}{lccccccc}
\toprule
& \multicolumn{2}{c}{Scoring \& Ranking} & \multicolumn{2}{c}{Docking} & \multicolumn{3}{c}{Screening} \\
\cmidrule(lr){2-3}\cmidrule(lr){4-5}\cmidrule(lr){6-8}
Methods
& $R_p$ & $R_s$
& {\renewcommand{\arraystretch}{0.2} 
   \begin{tabular}{@{}c@{}} 
     $SR_1$ \\ 
     (native poses \\ 
     excluded) 
   \end{tabular}}

& {\renewcommand{\arraystretch}{0.2}
   \begin{tabular}{@{}c@{}} 
     $SR_1$ \\ 
     (native poses \\ 
     included) 
   \end{tabular}}
& \multicolumn{2}{c}{\raisebox{2ex}{Forward}} 
& \raisebox{2ex}{Reverse} \\[-15pt]
\cmidrule(lr){6-7}

& & & & & $EF_{1\%}$ & $SR_{1\%}$ & $SR_{1\%}$ \\
\midrule
AutoDock Vina             & 0.604 & 0.528 & 0.846 & 0.902 &  7.70 & 0.298 & 0.137 \\
ChemPLP@GOLD                 & 0.614 & 0.633 & 0.832 & 0.860 & 11.91 & 0.351 & 0.165 \\
GlideScore-SP               & 0.513 & 0.419 & 0.846 & 0.877 & 11.44 & 0.368 & 0.175 \\
$\Delta_{\mathrm{vina}}\mathrm{RF}_{20}$  & 0.739 & 0.635 & 0.849 & 0.891 & 12.36 & 0.456 & ---   \\
OnionNet-SFCT+Vina          & 0.428 & 0.393 & ---   & 0.937 & 15.50 & 0.421 & ---   \\
PIGNet                      & 0.761 & 0.682 & ---   & 0.87  & 19.6  & 0.554 & ---   \\
PIGNet2                    & 0.747 & 0.651 & ---   & 0.930 & 24.90 & 0.667 & ---   \\
DeepDock                   & 0.460 & 0.425 & ---   & 0.891 & 16.40 & 0.439 & ---   \\
RTMScore                   & 0.456 & 0.539 & 0.93  & 0.961 & 25.63 & 0.596 & 0.361 \\
GenScore$_{\mathit{base}}$       & 0.494 & 0.543 & 0.921 & 0.952 & 22.59 & 0.52  & 0.3   \\
GenScore$_{\mathit{joint}}$     & 0.833 & 0.726 & 0.89  & 0.905 & 17.35 & 0.544 & 0.15  \\
GenScore$_{\mathit{ft}}$ & 0.824 & 0.709 & 0.926 & 0.953 & 26.41 & 0.684 & 0.205 \\
BioLM-Score$_{\mathit{base}}$  & 0.496 & 0.545 & 0.934 & \textbf{0.964} & 25.13 & 0.585 & \textbf{0.372} \\
BioLM-Score$_{\mathit{joint}}$ & \textbf{0.841} & \textbf{0.737} & 0.910 & 0.931 & 18.47 & 0.591 & 0.166 \\
BioLM-Score$_{\mathit{ft}}$    & 0.809 & 0.681 & \textbf{0.943} & 0.957 & \textbf{29.51} & \textbf{0.772} & 0.275 \\
\bottomrule
\end{tabular}}
\caption{BioLM-Score and representative scoring functions on the CASF-2016 benchmark. The best results are highlighted in \textbf{bold}.}
\label{tab:casf2016}
\end{table*}

\subsection{Pairwise Interaction Modeling and Geometric Scoring}
We first construct a residue-atom interaction embedding $\mathbf{z}_{uv} \in \mathbb{R}^{2d}$ for each pair by concatenating their respective features: $\mathbf{z}_{uv} = [\mathbf{Z}_P[u]\,\|\,\mathbf{Z}_L[v]]$.
To quantify geometric plausibility, we formulate the conditional distribution of the pairwise spatial distance $d_{uv}$ using a $K$-component one-dimensional MDN, where $d_{uv}$ represents the minimum Euclidean distance between ligand atom $v$ and any atom in residue $u$.
Specifically, we project $\mathbf{z}_{uv}$ via independent linear heads to obtain the mixture parameters:
\begin{equation}
    \begin{aligned}
\boldsymbol{\pi}_{uv} &= \mathrm{softmax}(\mathbf{W}_{\pi} \mathbf{z}_{uv}),\\
\boldsymbol{\mu}_{uv} &= \mathbf{W}_{\mu} \mathbf{z}_{uv},\\
\boldsymbol{\sigma}_{uv} &= \mathrm{softplus}(\mathbf{W}_{\sigma} \mathbf{z}_{uv}),
    \end{aligned}
\end{equation}
where $\boldsymbol{\pi}_{uv}, \boldsymbol{\mu}_{uv}, \boldsymbol{\sigma}_{uv} \in \mathbb{R}^{K}$ denote the mixing coefficients, means, and standard deviations, respectively.
The probability density of a given distance $d_{uv}$ is then given by:
\begin{equation}
p(d_{uv},|,\mathbf{z}_{uv}) = \sum_{k=1}^{K} \pi_{uv,k} \mathcal{N}\left(d_{uv}; \mu_{uv,k}, \sigma_{uv,k}^{2}\right).
\end{equation}
Finally, we formulate the pose score $s(P,L)$ as the aggregated log-likelihood over the set of interacting pairs $\mathcal{S}(L,P)$, defined as pairs with $d_{uv} \le 5\,\text{\AA}$ to enforce local geometric consistency:
\begin{equation}
s(P,L) = \sum_{(u,v)\in\mathcal{S}(L,P)} \log p(d_{uv} \,|\, \mathbf{z}_{uv}).
\end{equation}
Consequently, the final probabilistic geometric potential is defined as $E(P,L) = -s(P,L)$.

\subsection{Training Objectives}
\paragraph{Geometry Likelihood Loss.}
We minimize the negative log-likelihood loss over residue-atom pairs within a distance cutoff of $7\,\text{\AA}$:
\begin{equation}
    \mathcal{L}_{\mathrm{base}} = -\sum_{(u,v)\in\mathcal{T}(L,P)} \log p(d_{uv} \,|\,\mathbf{z}_{uv}),
\end{equation}
where $\mathcal{T}(L,P)=\{(u,v):d_{uv}\le 7\text{\AA}\}$ denotes the set of training pairs.
\paragraph{Auxiliary Chemical Supervision.}
To preserve chemical semantics, we employ auxiliary heads to reconstruct atom and bond types from their respective embeddings. We minimize the weighted sum of cross-entropy losses:
\begin{equation}
    \mathcal{L}_{\mathrm{aux}} = \lambda_{\mathrm{atom}}\mathcal{L}_{\mathrm{atom}} + \lambda_{\mathrm{bond}}\mathcal{L}_{\mathrm{bond}},
\end{equation}
where $\lambda_{\mathrm{atom}}$ and $\lambda_{\mathrm{bond}}$ are weighting hyperparameters.
\paragraph{Affinity Alignment.}
Acknowledging that geometric plausibility does not strictly imply binding affinity (as observed in \textbf{GenScore}), we introduce an optional affinity term $\mathcal{L}_{\text{aff}}$ to calibrate the scoring function without altering the underlying geometric potential definition:
\begin{equation}
    \mathcal{L}_{\text{aff}} = - \text{Corr}(\hat{s}, y_{\text{aff}}),
\end{equation}
where $\text{Corr}$ denotes the Pearson correlation coefficient between the predicted score $\hat{s}$ and the experimental affinity $y_{\text{aff}}$.
\paragraph{Training Strategies.}
Based on this framework, we instantiate three model variants to address different application scenarios, weighted by a coefficient $\alpha$:
\begin{itemize}
    \item \textbf{BioLM-Score$_{\text{base}}$ (Geometry-only):} The model is optimized exclusively via a geometric objective to learn a robust conformational potential:
    \begin{equation}
        \mathcal{L}^{(1)} = \mathcal{L}_{\text{base}}.
    \end{equation}
    \item \textbf{BioLM-Score$_{\text{joint}}$ (Joint Training):} We introduce affinity supervision simultaneously with geometric learning to enhance the model's ranking and screening capabilities:
\begin{equation}
\mathcal{L}^{(2)} = \mathcal{L}_{\text{base}} + \alpha\mathcal{L}_{\text{aff}}.
\end{equation}

\item \textbf{BioLM-Score$_{\text{ft}}$ (Two-stage Fine-tuning):} To prioritize geometric consistency, the model is first pre-trained with $\mathcal{L}^{(1)}$ and subsequently fine-tuned with the affinity term:
\begin{equation}
\mathcal{L}^{(3)}: \text{Stage-1 } \mathcal{L}_{\text{base}} \to \text{Stage-2 } \mathcal{L}_{\text{base}} + \alpha\mathcal{L}_{\text{aff}}.
\end{equation}
\end{itemize}

\begin{table*}[htbp]
\centering
\resizebox{\textwidth}{!}{%
\begin{tabular}{lcccccccccc}
\toprule
\multirow{2}{*}{Methods} & \multicolumn{2}{c}{AUROC} & \multicolumn{2}{c}{BEDROC($\alpha=80.5$)} & \multicolumn{2}{c}{$EF_{0.5\%}$} & \multicolumn{2}{c}{$EF_{1\%}$} & \multicolumn{2}{c}{$EF_{1\%}$} \\ 
\cmidrule(lr){2-3} \cmidrule(lr){4-5} \cmidrule(lr){6-7} \cmidrule(lr){8-9} \cmidrule(lr){10-11}
 & Mean & Median & Mean & Median & Mean & Median & Mean & Median & Mean & Median \\
\midrule
Glide SP & 0.747 & 0.754 & 0.385 & 0.314 & 14.610 & 13.300 & 12.470 & 9.610 & 6.300 & 5.970 \\
\midrule

RTMScore & 0.772 & 0.790 & 0.504 & 0.546 & 18.953 & 21.622 & 17.377 & 20.513 & 8.410 & 8.000 \\
\midrule

\multicolumn{11}{l}{\textit{Graph Transformer (GT)}} \\
GenScore(GT\_0) & 0.771 & 0.787 & 0.507 & \textbf{0.559} & 19.082 & 21.832 & 17.835 & 19.645 & 8.508 & 8.865 \\
GenScore(GT\_1.0) & 0.752 & 0.771 & 0.436 & 0.449 & 16.801 & 18.938 & 15.288 & 16.517 & 7.485 & 7.362 \\
GenScore(GT\_ft\_1.0) & 0.760 & 0.773 & 0.507 & 0.523 & 19.258 & 23.563 & 17.651 & 19.487 & 8.433 & 8.641 \\
BioLM-Score(GT\_0) & \textbf{0.774} & \textbf{0.794} & \textbf{0.515} & 0.550 & \textbf{19.716} & \textbf{24.074} & \textbf{18.269} & 19.982 & \textbf{8.576} & \textbf{8.883} \\
BioLM-Score(GT\_1.0) & 0.764 & 0.777 & 0.430 & 0.459 & 16.360 & 18.910 & 15.023 & 15.654 & 7.582 & 8.205 \\
BioLM-Score(GT\_ft\_1.0) & 0.772 & 0.778 & 0.509 & 0.538 & 19.388 & 22.758 & 17.893 & \textbf{20.693} & 8.485 & 8.474 \\
\midrule

\multicolumn{11}{l}{\textit{GatedGCN}} \\
GenScore(GatedGCN\_0) & 0.765 & 0.773 & 0.490 & \textbf{0.565} & 18.338 & 20.693 & 17.150 & 18.947 & 8.296 & \textbf{8.812} \\
GenScore(GatedGCN\_1.0) & 0.750 & 0.764 & 0.415 & 0.413 & 15.907 & 17.058 & 14.220 & 14.402 & 7.254 & 7.071 \\
GenScore(GatedGCN\_ft\_1.0) & 0.751 & 0.759 & 0.478 & 0.510 & 18.225 & 21.063 & 16.770 & 18.654 & 8.097 & 8.079 \\
BioLM-Score(GatedGCN\_0) & \textbf{0.767} & \textbf{0.783} & \textbf{0.508} & 0.530 & \textbf{19.361} & \textbf{23.830} & \textbf{18.079} & 19.681 & \textbf{8.414} & 7.947 \\
BioLM-Score(GatedGCN\_1.0) & 0.762 & 0.776 & 0.444 & 0.450 & 17.384 & 20.342 & 15.740 & 16.374 & 7.665 & 7.587 \\
BioLM-Score(GatedGCN\_ft\_1.0) & 0.755 & 0.774 & 0.504 & 0.543 & 19.268 & 21.432 & 18.039 & \textbf{20.513} & 8.213 & 8.209 \\
\bottomrule
\end{tabular}%
}
\caption{Screening performance comparison of different scoring methods on the DEKOIS 2.0. The best results are highlighted in \textbf{bold}.}
\label{DEKOIS}
\end{table*}

\section{Experiments}
\label{sec:experiments}

\subsection{Datasets and Preprocessing}
\label{sec:data_preprocess}

\paragraph{CASF-2016 Benchmark.}
For comprehensive evaluation, we employ the CASF-2016 benchmark, a widely recognized standard derived from the PDBbind core set. This dataset consists of 285 distinct protein-ligand complexes, structured into 57 target groups, with each group containing 5 co-crystal ligands. We utilize this benchmark to assess the model's capabilities across four standard tasks: scoring, ranking, docking, and screening.

\paragraph{Virtual Screening Benchmark.}
To evaluate the generalization capability of our scoring function in virtual screening (VS) scenarios, we employ the DEKOIS 2.0 dataset as an external benchmark. DEKOIS 2.0 encompasses 81 diverse targets. For each target, the dataset provides 30 known active ligands and 1,200 decoys, forming a challenging candidate library composed of actives and decoys. To simulate a realistic screening pipeline, we first utilize Glide SP to generate docking poses for each candidate compound, retaining a maximum of 10 poses per compound. Subsequently, our proposed model (along with baseline methods) is applied to rescore these pre-generated poses. Finally, compounds are ranked within each target based on the rescored values to assess the virtual screening performance.

\subsection{Performance on CASF-2016}
\label{sec:results_casf}

As shown in Table~\ref{tab:casf2016}, BioLM-Score achieves state-of-the-art performance across all four evaluation metrics (scoring, ranking, docking, and screening), effectively addressing the performance trade-offs commonly observed in existing methods (for detailed results, see Supplementary Table~1). 

As shown in Table~\ref{tab:casf2016}, BioLM-Score achieves state-of-the-art performance across all four metrics (scoring, ranking, docking, and screening), effectively overcoming the trade-offs inherent in existing methods. Traditional physics-based functions (e.g., AutoDock Vina) excel in docking via explicit physical modeling but struggle with scoring and screening due to oversimplified solvation and entropy treatments. While hybrid methods (e.g., $\Delta_{\mathrm{vina}}\mathrm{RF}_{20}$) calibrate these terms using machine learning, they remain limited by the expressivity of physical descriptors. Conversely, MDN-based geometric methods (e.g., RTMScore) achieve robust docking success but often fail to map geometric likelihoods to thermodynamic affinity. Although GenScore addresses this by introducing an affinity-oriented objective, its GNN-based architecture captures only local topology, lacking global biological context. BioLM-Score distinguishes itself by redefining input representation: by integrating protein and ligand LMs, it incorporates evolutionary priors and global chemical semantics into geometric graphs. This enables the identification of true binders based on both geometric fit and biochemical validity, thus bridging the geometry-affinity gap and delivering superior ranking and screening without compromising docking accuracy.

\subsection{Evaluation on DEKOIS 2.0} \label{sec:results_dekois} Table~\ref{DEKOIS} presents the performance evaluation on the DEKOIS 2.0 benchmark. Glide SP is selected as the primary baseline, as it was utilized for decoy generation in this dataset. Compared to traditional scoring functions, distance-likelihood-based generative models (i.e., RTMScore, GenScore, and BioLM-Score) demonstrate superior performance, verifying the efficacy of learning interaction patterns directly from native structures.

Notably, BioLM-Score achieves a substantial margin over GenScore. With the same GatedGCN or GT backbones, BioLM-Score establishes a new state-of-the-art across nearly all metrics. We attribute this performance gap to the distinct conditioning of interaction probabilities. Unlike GenScore, which relies solely on local chemical environments that are often prone to noise within complex binding pockets, BioLM-Score leverages language model representations to enrich geometric features with semantic context. Specifically, language model embeddings introduce functional context to refine distance predictions, assigning higher probabilities to biologically and evolutionarily consistent interactions.

Furthermore, a comparative analysis underscores the critical role of edge-aware information routing. Specifically, the Structure-aware GT consistently outperforms GatedGCN baselines. Unlike the scalar sigmoid gating in GatedGCN, GT employs multi-head attention explicitly modulated by edge features via query-key interactions. This mechanism facilitates the joint optimization of node-edge dynamics, enabling the capture of complex structural dependencies that scalar gating misses.

\subsection{Ablation Study on Feature Fusion and Training Protocols}

We conduct an ablation study using the CASF-2016 benchmark to investigate optimal strategies for fusing linguistic and geometric features and to analyze the impact of affinity supervision on structural representation learning. Quantitative results are summarized in Table~\ref{fusion}, with detailed results presented in {Supplementary Table~2}.

\paragraph{Impact of Training Protocols.} We first analyze the interplay between structural modeling and affinity prediction. In the unsupervised setting ($\alpha=0$), the model focuses exclusively on geometric constraints, thereby achieving competitive performance in docking and screening tasks. However, the lack of explicit binding affinity supervision leads to suboptimal scoring correlations ($R_p$). Conversely, incorporating affinity loss ($\alpha=1$) significantly boosts scoring accuracy but introduces a trade-off, degrading docking success rates ($SR_1$) and screening efficiency. This suggests that a joint multi-task learning objective induces optimization conflicts between the generative and regression tasks. Notably, the proposed fine-tuning strategy effectively mitigates this conflict. By decoupling geometric pre-training from affinity refinement, this protocol achieves a superior balance: it retains the high structural precision of the unsupervised phase (comparable $SR_1$ to $\alpha=0$) while achieving state-of-the-art scoring accuracy (comparable $R_p$ to $\alpha=1$).

\paragraph{Effectiveness of Feature Fusion Architectures.} We examine four distinct fusion mechanisms for integrating language and GNN features. Contrary to expectations, the complex Cross-Attention (\texttt{cross\_attn}) strategy fails to translate into performance gains. In the fine-tuning setting, it underperforms simpler methods in scoring accuracy. We hypothesize that given the limited training data scale ($\sim$17k complexes), the over-parameterized attention mechanism suffers from optimization difficulties, resulting in poor generalization on the binding affinity regression task. Similarly, element-wise aggregation methods such as \texttt{add} and \texttt{gated\_add} lack the capacity to model high-order feature interactions, limiting screening performance. In contrast, the concatenation strategy (\texttt{concat}) consistently demonstrates superior performance. It achieves the optimal balance, particularly in the fine-tuning stage (highest $SR_1$ and competitive $R_p$). This suggests that preserving distinct feature representations from both modalities, without enforcing premature mixing, is the most effective strategy for this multimodal objective.

\subsection{Ablation Study on Biological Language Modalities}

To evaluate the individual contributions of sequence-based representations to geometric interaction modeling, we conducted ablation studies by selectively masking the Protein LM or Ligand LM components. Results across different training protocols are summarized in Table~\ref{llm}. More detailed results are provided in {Supplementary Table~3}.

\paragraph{Pivotal Role of Protein Semantics.} Consistently across all settings, removing the Protein LM causes the most significant performance degradation, particularly in docking and virtual screening tasks. For instance, under the fine-tuning protocol, eliminating protein language features results in a substantial decline in docking success rates ($SR_1$ decreases from 0.943 to 0.891) and screening efficiency ($EF_{1\%}$ from 29.51 to 23.61). This suggests that the geometric encoder (GNN) alone, while capturing local atomic interactions, fails to adequately encode the global functional context of the protein pocket. The Protein LM compensates for this deficiency by providing evolutionary and semantic information embedded in the amino acid sequence, which is crucial for distinguishing native binding poses from decoys.

\paragraph{Complementary Benefits of Ligand LM.}
The Ligand LM offers primarily complementary benefits. Ablating this module achieves moderate yet consistent degradation in docking ($SR_1$) and ranking ($R_s$), indicating that ligand language priors enhance multi-task performance without dominating geometric feasibility. Notably, the ablation reveals a specific trade-off: while enrichment efficiency ($EF_{1\%}$) remains robust (29.83 vs. 29.51), screening success rates ($SR_{1\%}$) and affinity correlations decline. This suggests that ligand semantics are essential for rebalancing optimization objectives, ensuring the model does not over-optimize specific geometric metrics at the expense of comprehensive scoring and ranking accuracy.

\paragraph{Synergy in the Full Model.} The full model consistently achieves the most balanced performance, particularly under the fine-tuning strategy. By effectively integrating global protein semantics and ligand chemical contexts with geometric features, the model demonstrates that a holistic multimodal approach is essential for state-of-the-art performance in both binding conformation selection and affinity ranking.

\begin{table}[t!]
  \centering
  \small
  \setlength{\tabcolsep}{3pt}   
  \renewcommand{\arraystretch}{1.0} 
  \begin{tabular}{llccccc}
    \toprule
    \multirow{2}{*}{Setting} & \multirow{2}{*}{Method} & \multicolumn{2}{c}{Scoring Ranking} & Docking & \multicolumn{2}{c}{Screening} \\
    \cmidrule(lr){3-4} \cmidrule(lr){5-5} \cmidrule(lr){6-7}
     & & $R_p$ & $R_s$ & $SR_1$ & $EF_{1\%}$ & $SR_{1\%}$ \\
    \midrule
    \multirow{4}{*}{$\alpha=0$} 
      & add        & 0.487 & 0.554 & 0.923 & 25.12 & 0.561 \\
      & gated\_add & 0.465 & 0.546 & 0.923 & 24.47 & 0.566 \\
      & cross\_attn& 0.473 & 0.560 & 0.933 & 23.08 & 0.579 \\
      & concat     & 0.496 & 0.545 & \textbf{0.934} & \textbf{25.13} & \textbf{0.585} \\
    \midrule
    \multirow{4}{*}{$\alpha=1$} 
      & add        & 0.821 & 0.732 & 0.902 & 17.53 & 0.552 \\
      & gated\_add & 0.827 & 0.739 & \textbf{0.916} & 18.32 & \textbf{0.667} \\
      & cross\_attn& 0.833 & 0.735 & 0.891 & 16.46 & 0.509 \\
      & concat     & \textbf{0.841} & \textbf{0.737} & 0.910 & \textbf{18.47} & 0.591 \\
    \midrule
    \multirow{4}{*}{finetune} 
      & add        & \textbf{0.825} & 0.698 & 0.937 & 26.88 & 0.737 \\
      & gated\_add & 0.809 & 0.688 & 0.930 & 29.04 & 0.772 \\
      & cross\_attn& 0.797 & \textbf{0.711} & 0.933 & 29.01 & 0.772 \\
      & concat     & 0.809 & 0.681 & \textbf{0.943} & \textbf{29.51} & \textbf{0.772} \\
    \bottomrule
  \end{tabular}
    \caption{Comparison of results of different feature fusion and training methods on the CASF-2016 benchmark}
  \label{fusion}
\end{table}

\begin{table}[t!]
  \centering
  \small
  \label{tab:ablation_study}
  \begin{tabular}{lccccc}
    \toprule
    \multirow{2}{*}{Variants} & Scoring & Ranking & Docking & \multicolumn{2}{c}{Screening} \\
    \cmidrule(lr){2-2} \cmidrule(lr){3-3} \cmidrule(lr){4-4} \cmidrule(lr){5-6}
     & $R_p$ & $R_s$ & $SR_1$ & $EF_{1\%}$ & $SR_{1\%}$ \\
    \midrule
    W/O protein & \textbf{0.809} & 0.660 & 0.891 & 23.61 & 0.569 \\
    W/O ligand  & \underline{0.806} & \underline{0.674} & \underline{0.926} & \textbf{29.83} & \underline{0.737} \\
    Full model  & \textbf{0.809} & \textbf{0.681} & \textbf{0.943} & \underline{29.51} & \textbf{0.772} \\
    \bottomrule
  \end{tabular}
    \caption{Ablation study of different variants on the benchmark. The best results are highlighted in \textbf{bold}, and the second-best results are \underline{underlined}.}
  \label{llm}
\end{table}

\subsection{Docking Performance}

To evaluate the discriminative capability of BioLM-Score for molecular docking, we integrate it into a search framework named \textbf{BSDock}. Conventional docking protocols typically couple a scoring function with a stochastic search algorithm. Adhering to this design, BSDock utilizes BioLM-Score as the primary objective function to guide a Differential Evolution (DE) based global search ~\cite{qin2008differential}.

While BioLM-Score effectively captures intermolecular protein-ligand interactions, it lacks explicit penalties for steric clashes or intramolecular geometric distortions. To ensure physical plausibility and mitigate structural anomalies, we incorporate a gradient-based local refinement step utilizing the AutoDock Vina scoring function. This hybrid strategy combines BioLM-Score for global pose optimization with a physics-inspired scoring function for structural regularization.

\begin{figure}[t!] 
    \centering
    \includegraphics[width=1.0\linewidth]{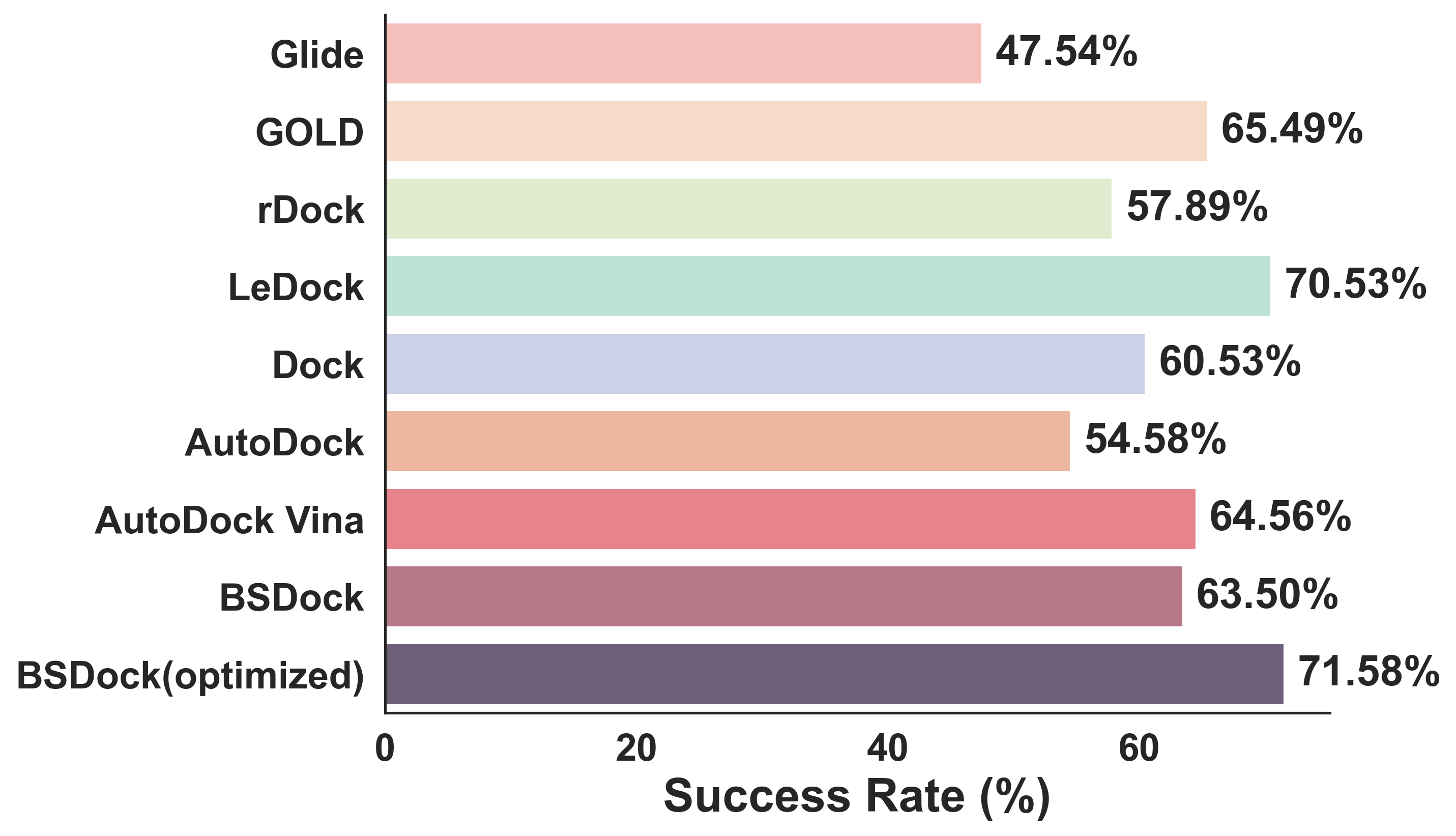} 
    
    \caption{Comparison of Docking Success Rates on CASF-2016. Success rate is defined as the percentage of top-1 ranked poses with an RMSD $< 2.0$ \AA\ relative to the crystal structure. Baselines include widely adopted commercial and academic software.}
    \label{tab:docking_success}
    
    \label{fig:dock}
\end{figure}

Figure~\ref{fig:dock} reports the performance on the CASF-2016 benchmark. Notably, even without local refinement, BSDock achieves a success rate of 63.5\%, comparable to established baselines such as AutoDock and rDock~\cite{ruiz2014rdock}. Incorporating the local refinement module, BSDock (optimized) attains a leading success rate, surpassing LeDock~\cite{liu2019using} and significantly outperforming the widely used AutoDock Vina. These results demonstrate that BioLM-Score provides effective guidance for conformational search, with performance further enhanced by physical constraint enforcement.

\section{Conclusion}

In this study, we have presented BioLM-Score, a novel protein–ligand scoring framework that bridges the gap between geometric deep learning and biomolecular representation learning. By integrating structure-aware geometric encoders with pre-trained biomolecular language models, our approach effectively captures both the local physico-chemical interactions defined by 3D graphs and the global semantic contexts encoded within protein sequences and ligand SMILES. This dual-branch architecture, coupled with a mixture density network, enables the prediction of multimodal interatomic distance distributions, establishing a scoring function that is both statistically grounded and physically interpretable.

Extensive empirical evaluations on the CASF-2016 and DEKOIS 2.0 benchmarks demonstrate that BioLM-Score achieves state-of-the-art performance across a broad spectrum of tasks, including scoring, ranking, docking, and virtual screening. Notably, our results highlight that incorporating evolutionary and chemical language priors significantly enhances the model's ability to generalize across diverse targets compared to purely structure-based baselines. Furthermore, we have established the practical utility of BioLM-Score as a differentiable optimization objective, proving its effectiveness in guiding conformational search and refining binding poses within docking protocols.

\newpage
\bibliographystyle{named}
\bibliography{ijcai26}

\begin{thebibliography}{}

\bibitem[\protect\citeauthoryear{Aldeghi \bgroup \em et al.\egroup }{2016}]{aldeghi2016accurate}
Matteo Aldeghi, Alexander Heifetz, Michael~J Bodkin, Stefan Knapp, and Philip~C Biggin.
\newblock Accurate calculation of the absolute free energy of binding for drug molecules.
\newblock {\em Chemical science}, 7(1):207--218, 2016.

\bibitem[\protect\citeauthoryear{Ban \bgroup \em et al.\egroup }{2017}]{ban2017best}
Fuqiang Ban, Kush Dalal, Huifang Li, Eric LeBlanc, Paul~S Rennie, and Artem Cherkasov.
\newblock Best practices of computer-aided drug discovery: lessons learned from the development of a preclinical candidate for prostate cancer with a new mechanism of action.
\newblock {\em Journal of chemical information and modeling}, 57(5):1018--1028, 2017.

\bibitem[\protect\citeauthoryear{Bauer \bgroup \em et al.\egroup }{2013}]{bauer2013evaluation}
Matthias~R Bauer, Tamer~M Ibrahim, Simon~M Vogel, and Frank~M Boeckler.
\newblock Evaluation and optimization of virtual screening workflows with dekois 2.0--a public library of challenging docking benchmark sets.
\newblock {\em Journal of chemical information and modeling}, 53(6):1447--1462, 2013.

\bibitem[\protect\citeauthoryear{Bresson and Laurent}{2017}]{bresson2017residual}
Xavier Bresson and Thomas Laurent.
\newblock Residual gated graph convnets.
\newblock {\em arXiv preprint arXiv:1711.07553}, 2017.

\bibitem[\protect\citeauthoryear{Chauhan and Singh}{2018}]{chauhan2018review}
Nitin~Kumar Chauhan and Krishna Singh.
\newblock A review on conventional machine learning vs deep learning.
\newblock In {\em 2018 International conference on computing, power and communication technologies (GUCON)}, pages 347--352. IEEE, 2018.

\bibitem[\protect\citeauthoryear{Chen \bgroup \em et al.\egroup }{2025}]{chen2025hitscreen}
Geng Chen, Jinbiao Liao, Yanzhen Yu, Kaixin Le, Hui Zhao, Yiyang Qin, Lvtao Cai, and Rong Sheng.
\newblock Hitscreen: A sequence-based drug virtual screening approach using data augmentation and protein language models.
\newblock {\em Journal of Chemical Information and Modeling}, 65(19):10152--10166, 2025.

\bibitem[\protect\citeauthoryear{Drews}{2000}]{drews2000drug}
Jurgen Drews.
\newblock Drug discovery: a historical perspective.
\newblock {\em science}, 287(5460):1960--1964, 2000.

\bibitem[\protect\citeauthoryear{Dwivedi and Bresson}{2020}]{dwivedi2020generalization}
Vijay~Prakash Dwivedi and Xavier Bresson.
\newblock A generalization of transformer networks to graphs.
\newblock {\em arXiv preprint arXiv:2012.09699}, 2020.

\bibitem[\protect\citeauthoryear{{ESM Team and others}}{2024}]{esm2024cambrian}
{ESM Team and others}.
\newblock {ESM Cambrian}: Revealing the mysteries of proteins with unsupervised learning.
\newblock \url{https://www.evolutionaryscale.ai/blog/esm-cambrian}, 2024.
\newblock Evolutionary Scale Blog. Accessed: 2024-01-17.

\bibitem[\protect\citeauthoryear{Friesner \bgroup \em et al.\egroup }{2006}]{glide}
Richard~A Friesner, Robert~B Murphy, Matthew~P Repasky, Leah~L Frye, Jeremy~R Greenwood, Thomas~A Halgren, Paul~C Sanschagrin, and Daniel~T Mainz.
\newblock Extra precision glide: Docking and scoring incorporating a model of hydrophobic enclosure for protein- ligand complexes.
\newblock {\em Journal of medicinal chemistry}, 49(21):6177--6196, 2006.

\bibitem[\protect\citeauthoryear{Gorgulla \bgroup \em et al.\egroup }{2020}]{gorgulla2020open}
Christoph Gorgulla, Andras Boeszoermenyi, Zi-Fu Wang, Patrick~D Fischer, Paul~W Coote, Krishna~M Padmanabha~Das, Yehor~S Malets, Dmytro~S Radchenko, Yurii~S Moroz, David~A Scott, et~al.
\newblock An open-source drug discovery platform enables ultra-large virtual screens.
\newblock {\em Nature}, 580(7805):663--668, 2020.

\bibitem[\protect\citeauthoryear{Irwin \bgroup \em et al.\egroup }{2022}]{irwin2022chemformer}
Ross Irwin, Spyridon Dimitriadis, Jiazhen He, and Esben~Jannik Bjerrum.
\newblock Chemformer: a pre-trained transformer for computational chemistry.
\newblock {\em Machine Learning: Science and Technology}, 3(1):015022, 2022.

\bibitem[\protect\citeauthoryear{Jin \bgroup \em et al.\egroup }{2023}]{jin2023capla}
Zhi Jin, Tingfang Wu, Taoning Chen, Deng Pan, Xuejiao Wang, Jingxin Xie, Lijun Quan, and Qiang Lyu.
\newblock Capla: improved prediction of protein--ligand binding affinity by a deep learning approach based on a cross-attention mechanism.
\newblock {\em Bioinformatics}, 39(2):btad049, 2023.

\bibitem[\protect\citeauthoryear{Lam \bgroup \em et al.\egroup }{2024}]{lam2024protein}
Hilbert Yuen~In Lam, Jia~Sheng Guan, Xing~Er Ong, Robbe Pincket, and Yuguang Mu.
\newblock Protein language models are performant in structure-free virtual screening.
\newblock {\em Briefings in Bioinformatics}, 25(6):bbae480, 2024.

\bibitem[\protect\citeauthoryear{Liu and Xu}{2019}]{liu2019using}
Ni~Liu and Zhibin Xu.
\newblock Using ledock as a docking tool for computational drug design.
\newblock In {\em IOP Conference Series: Earth and Environmental Science}, volume 218, page 012143. IOP Publishing, 2019.

\bibitem[\protect\citeauthoryear{Lu \bgroup \em et al.\egroup }{2019}]{vinaxgb}
Jianing Lu, Xuben Hou, Cheng Wang, and Yingkai Zhang.
\newblock Incorporating explicit water molecules and ligand conformation stability in machine-learning scoring functions.
\newblock {\em Journal of chemical information and modeling}, 59(11):4540--4549, 2019.

\bibitem[\protect\citeauthoryear{Luo \bgroup \em et al.\egroup }{2024}]{luo2024enhancing}
Ding Luo, Dandan Liu, Xiaoyang Qu, Lina Dong, and Binju Wang.
\newblock Enhancing generalizability in protein--ligand binding affinity prediction with multimodal contrastive learning.
\newblock {\em Journal of chemical information and modeling}, 64(6):1892--1906, 2024.

\bibitem[\protect\citeauthoryear{Mayr and Bojanic}{2009}]{mayr2009novel}
Lorenz~M Mayr and Dejan Bojanic.
\newblock Novel trends in high-throughput screening.
\newblock {\em Current opinion in pharmacology}, 9(5):580--588, 2009.

\bibitem[\protect\citeauthoryear{Meli \bgroup \em et al.\egroup }{2021}]{deltaae}
Rocco Meli, Andrew Anighoro, Mike~J Bodkin, Garrett~M Morris, and Philip~C Biggin.
\newblock Learning protein-ligand binding affinity with atomic environment vectors.
\newblock {\em Journal of Cheminformatics}, 13(1):59, 2021.

\bibitem[\protect\citeauthoryear{Meli \bgroup \em et al.\egroup }{2022}]{meli2022scoring}
Rocco Meli, Garrett~M Morris, and Philip~C Biggin.
\newblock Scoring functions for protein-ligand binding affinity prediction using structure-based deep learning: a review.
\newblock {\em Frontiers in bioinformatics}, 2:885983, 2022.

\bibitem[\protect\citeauthoryear{M{\'e}ndez-Lucio \bgroup \em et al.\egroup }{2021}]{deepdock}
Oscar M{\'e}ndez-Lucio, Mazen Ahmad, Ehecatl~Antonio del Rio-Chanona, and J{\"o}rg~Kurt Wegner.
\newblock A geometric deep learning approach to predict binding conformations of bioactive molecules.
\newblock {\em Nature Machine Intelligence}, 3(12):1033--1039, 2021.

\bibitem[\protect\citeauthoryear{Monticelli and Tieleman}{2012}]{monticelli2012force}
Luca Monticelli and D~Peter Tieleman.
\newblock Force fields for classical molecular dynamics.
\newblock {\em Biomolecular simulations: Methods and protocols}, pages 197--213, 2012.

\bibitem[\protect\citeauthoryear{Moon \bgroup \em et al.\egroup }{2022}]{pignet}
Seokhyun Moon, Wonho Zhung, Soojung Yang, Jaechang Lim, and Woo~Youn Kim.
\newblock Pignet: a physics-informed deep learning model toward generalized drug--target interaction predictions.
\newblock {\em Chemical Science}, 13(13):3661--3673, 2022.

\bibitem[\protect\citeauthoryear{Moon \bgroup \em et al.\egroup }{2024}]{pignet2}
Seokhyun Moon, Sang-Yeon Hwang, Jaechang Lim, and Woo~Youn Kim.
\newblock Pignet2: a versatile deep learning-based protein--ligand interaction prediction model for binding affinity scoring and virtual screening.
\newblock {\em Digital Discovery}, 3(2):287--299, 2024.

\bibitem[\protect\citeauthoryear{Morris \bgroup \em et al.\egroup }{2009}]{morris2009autodock4}
Garrett~M Morris, Ruth Huey, William Lindstrom, Michel~F Sanner, Richard~K Belew, David~S Goodsell, and Arthur~J Olson.
\newblock Autodock4 and autodocktools4: Automated docking with selective receptor flexibility.
\newblock {\em Journal of computational chemistry}, 30(16):2785--2791, 2009.

\bibitem[\protect\citeauthoryear{Qin \bgroup \em et al.\egroup }{2008}]{qin2008differential}
A~Kai Qin, Vicky~Ling Huang, and Ponnuthurai~N Suganthan.
\newblock Differential evolution algorithm with strategy adaptation for global numerical optimization.
\newblock {\em IEEE transactions on Evolutionary Computation}, 13(2):398--417, 2008.

\bibitem[\protect\citeauthoryear{Ruiz-Carmona \bgroup \em et al.\egroup }{2014}]{ruiz2014rdock}
Sergio Ruiz-Carmona, Daniel Alvarez-Garcia, Nicolas Foloppe, A~Beatriz Garmendia-Doval, Szilveszter Juhos, Peter Schmidtke, Xavier Barril, Roderick~E Hubbard, and S~David Morley.
\newblock rdock: a fast, versatile and open source program for docking ligands to proteins and nucleic acids.
\newblock {\em PLoS computational biology}, 10(4):e1003571, 2014.

\bibitem[\protect\citeauthoryear{Shen \bgroup \em et al.\egroup }{2021}]{shen2021beware}
Chao Shen, Ye~Hu, Zhe Wang, Xujun Zhang, Jinping Pang, Gaoang Wang, Haiyang Zhong, Lei Xu, Dongsheng Cao, and Tingjun Hou.
\newblock Beware of the generic machine learning-based scoring functions in structure-based virtual screening.
\newblock {\em Briefings in Bioinformatics}, 22(3):bbaa070, 2021.

\bibitem[\protect\citeauthoryear{Shen \bgroup \em et al.\egroup }{2022}]{rtmscore}
Chao Shen, Xujun Zhang, Yafeng Deng, Junbo Gao, Dong Wang, Lei Xu, Peichen Pan, Tingjun Hou, and Yu~Kang.
\newblock Boosting protein--ligand binding pose prediction and virtual screening based on residue--atom distance likelihood potential and graph transformer.
\newblock {\em Journal of Medicinal Chemistry}, 65(15):10691--10706, 2022.

\bibitem[\protect\citeauthoryear{Shen \bgroup \em et al.\egroup }{2023}]{genscore}
Chao Shen, Xujun Zhang, Chang-Yu Hsieh, Yafeng Deng, Dong Wang, Lei Xu, Jian Wu, Dan Li, Yu~Kang, Tingjun Hou, et~al.
\newblock A generalized protein--ligand scoring framework with balanced scoring, docking, ranking and screening powers.
\newblock {\em Chemical Science}, 14(30):8129--8146, 2023.

\bibitem[\protect\citeauthoryear{Song \bgroup \em et al.\egroup }{2021}]{song2021protein}
Shuangbao Song, Xingqian Chen, Yanxin Zhang, Zheng Tang, and Yuki Todo.
\newblock Protein--ligand docking using differential evolution with an adaptive mechanism.
\newblock {\em Knowledge-Based Systems}, 231:107433, 2021.

\bibitem[\protect\citeauthoryear{Su \bgroup \em et al.\egroup }{2018}]{su2018comparative}
Minyi Su, Qifan Yang, Yu~Du, Guoqin Feng, Zhihai Liu, Yan Li, and Renxiao Wang.
\newblock Comparative assessment of scoring functions: the casf-2016 update.
\newblock {\em Journal of chemical information and modeling}, 59(2):895--913, 2018.

\bibitem[\protect\citeauthoryear{Trott and Olson}{2010}]{vina}
Oleg Trott and Arthur~J Olson.
\newblock Autodock vina: improving the speed and accuracy of docking with a new scoring function, efficient optimization, and multithreading.
\newblock {\em Journal of computational chemistry}, 31(2):455--461, 2010.

\bibitem[\protect\citeauthoryear{Verdonk \bgroup \em et al.\egroup }{2003}]{gold}
Marcel~L Verdonk, Jason~C Cole, Michael~J Hartshorn, Christopher~W Murray, and Richard~D Taylor.
\newblock Improved protein--ligand docking using gold.
\newblock {\em Proteins: Structure, Function, and Bioinformatics}, 52(4):609--623, 2003.

\bibitem[\protect\citeauthoryear{Wang and Zhang}{2017}]{vinarf}
Cheng Wang and Yingkai Zhang.
\newblock Improving scoring-docking-screening powers of protein--ligand scoring functions using random forest.
\newblock {\em Journal of computational chemistry}, 38(3):169--177, 2017.

\bibitem[\protect\citeauthoryear{Wang \bgroup \em et al.\egroup }{2002}]{x-score}
Renxiao Wang, Luhua Lai, and Shaomeng Wang.
\newblock Further development and validation of empirical scoring functions for structure-based binding affinity prediction.
\newblock {\em Journal of computer-aided molecular design}, 16(1):11--26, 2002.

\bibitem[\protect\citeauthoryear{Wang \bgroup \em et al.\egroup }{2024}]{igmodel}
Zechen Wang, Sheng Wang, Yangyang Li, Jingjing Guo, Yanjie Wei, Yuguang Mu, Liangzhen Zheng, and Weifeng Li.
\newblock A new paradigm for applying deep learning to protein--ligand interaction prediction.
\newblock {\em Briefings in bioinformatics}, 25(3):bbae145, 2024.

\bibitem[\protect\citeauthoryear{Wu \bgroup \em et al.\egroup }{2018}]{wu2018coach}
Qi~Wu, Zhenling Peng, Yang Zhang, and Jianyi Yang.
\newblock Coach-d: improved protein--ligand binding sites prediction with refined ligand-binding poses through molecular docking.
\newblock {\em Nucleic acids research}, 46(W1):W438--W442, 2018.

\bibitem[\protect\citeauthoryear{Xia \bgroup \em et al.\egroup }{2025}]{xia2025normalized}
Song Xia, Yaowen Gu, and Yingkai Zhang.
\newblock Normalized protein--ligand distance likelihood score for end-to-end blind docking and virtual screening.
\newblock {\em Journal of Chemical Information and Modeling}, 65(3):1101--1114, 2025.

\bibitem[\protect\citeauthoryear{Yang \bgroup \em et al.\egroup }{2022}]{yang2022protein}
Chao Yang, Eric~Anthony Chen, and Yingkai Zhang.
\newblock Protein--ligand docking in the machine-learning era.
\newblock {\em Molecules}, 27(14):4568, 2022.

\bibitem[\protect\citeauthoryear{Zhang \bgroup \em et al.\egroup }{2024}]{zhang2024advancing}
Xujun Zhang, Chao Shen, Haotian Zhang, Yu~Kang, Chang-Yu Hsieh, and Tingjun Hou.
\newblock Advancing ligand docking through deep learning: challenges and prospects in virtual screening.
\newblock {\em Accounts of chemical research}, 57(10):1500--1509, 2024.

\bibitem[\protect\citeauthoryear{Zhang \bgroup \em et al.\egroup }{2025}]{zhang2025graph}
Odin Zhang, Haitao Lin, Xujun Zhang, Xiaorui Wang, Zhenxing Wu, Qing Ye, Weibo Zhao, Jike Wang, Kejun Ying, Yu~Kang, et~al.
\newblock Graph neural networks in modern ai-aided drug discovery.
\newblock {\em Chemical Reviews}, 125(20):10001--10103, 2025.

\bibitem[\protect\citeauthoryear{Zheng \bgroup \em et al.\egroup }{2022}]{onionnet}
Liangzhen Zheng, Jintao Meng, Kai Jiang, Haidong Lan, Zechen Wang, Mingzhi Lin, Weifeng Li, Hongwei Guo, Yanjie Wei, and Yuguang Mu.
\newblock Improving protein--ligand docking and screening accuracies by incorporating a scoring function correction term.
\newblock {\em Briefings in Bioinformatics}, 23(3):bbac051, 2022.

\end{thebibliography}

\end{document}